\documentclass{article}
\usepackage[left=1.5in,right=1.5in,top=1.5in,bottom=1.5in]{geometry}

\usepackage{listings}
\usepackage[hang,flushmargin]{footmisc}
\usepackage[pdftex]{graphicx}
  \DeclareGraphicsExtensions{.pdf,.jpeg,.png}
\usepackage[caption=false,font=footnotesize]{subfig}
\usepackage{authblk}

\begin{document}

\title{Alchemist: An Apache Spark $\Leftrightarrow$ MPI Interface\footnote{Accepted for publication in Concurrency and Computation: Practice and Experience, Special Issue on the Cray User Group 2018}}

\author[1]{Alex Gittens\thanks{{\tt gittea@rpi.edu}}}
\author[2]{Kai Rothauge\thanks{{\tt kai.rothauge@berkeley.edu}}}
\author[2]{Shusen Wang}
\author[2]{Michael W. Mahoney}
\author[2]{Jey Kottalam}
\author[3]{Lisa Gerhardt}
\author[3]{Prabhat}
\author[4]{Michael Ringenburg}
\author[4]{Kristyn Maschhoff}
\affil[1]{Rensselaer Polytechnic Institute, Troy, NY}
\affil[2]{UC Berkeley, Berkeley, CA}
\affil[3]{NERSC/LBNL, Berkeley, CA}
\affil[4]{Cray Inc., Seattle, WA}

\maketitle

\begin{abstract}
The Apache Spark framework for distributed computation is popular in the data analytics community due to its ease of use, but its MapReduce-style programming model can incur significant overheads when performing computations that do not map directly onto this model. 
One way to mitigate these costs is to off-load computations onto MPI codes.
In recent work, we introduced Alchemist, a system for the analysis of large-scale data sets.
Alchemist calls MPI-based libraries from within Spark applications, and it has minimal coding, communication, and memory overheads. 
In particular, Alchemist allows users to retain the productivity benefits of working within the Spark software ecosystem without sacrificing performance efficiency in linear algebra, machine learning, and other related computations. 

In this paper, we discuss the motivation behind the development of Alchemist, and we provide a detailed overview its design and usage. 
We also demonstrate the efficiency of our approach on medium-to-large data sets, using some standard linear algebra operations, namely matrix multiplication and the truncated singular value decomposition of a dense matrix, and we compare the performance of Spark with that of Spark+Alchemist. 
These computations are run on the NERSC supercomputer Cori Phase 1, a Cray XC40.
\end{abstract}

\section{Introduction}
\label{sec:introduction}

\textit{Alchemist}, a framework for interfacing Apache Spark applications with MPI-based codes, was recently introduced in~\cite{Gittens2018a}. In this paper, we take a closer look at the motivation, design, and usage aspects of Alchemist.

\subsection{Motivation}

Apache Spark is a distributed computation framework that was introduced to facilitate processing and analyzing the huge amount of data generated over a wide range of applications~\cite{Zaharia2016}. 
It provides \emph{high productivity computing} interfaces for the data science community, and has seen substantial progress in its development and adoption in recent years. The \emph{high performance computing} (HPC) community, however, has so far been slow to embrace Spark, due in part to its low performance. 
While Spark performs well on certain types of data analysis, a recent case study illustrated a stark difference in computing times when performing some common numerical linear algebra computations in Apache Spark, compared to using MPI-based routines written in C or C++ (C+MPI)~\cite{Gittens2016}. 

One of the matrix factorizations considered in that study was the truncated singular-value decomposition (SVD). This is a ubiquitous matrix factorization, used in fields such as climate modeling, genetics neuroscience, and mathematical finance (among many others), but it is particularly challenging for Spark because the iterative nature of SVD algorithms leads to substantial communication and synchronization overheads. The results of an extensive and detailed evaluation (see Figures 5 and 6 in~\cite{Gittens2016}) show not only that Spark is more than an order of magnitude slower than the equivalent procedure implemented using C+MPI for datasets in the 10TB size range, but also that Spark's overheads in fact dominate and anti-scale, i.e., the overheads take up an increasing amount of the computational workload relative to the actual matrix factorization calculations as the number of nodes used increases~\cite{Gittens2016}.

Spark follows a data parallel, bulk synchronous programming model, where a single driver process manages the execution of the submitted applications, which are then carried out on executor processes. 
The sole distributed data type for Spark is the Resilient Distributed Dataset (RDD), an immutable distributed array that is partitioned across the executors. 
Each application submitted to a Spark cluster is converted to a computational graph on the driver, and this graph is then compiled into a sequence of stages, collections of computational tasks that are executed simultaneously on each partition of the RDDs. 
The tasks in each stage run on the executor that holds the relevant partition, with inter-process communication happening only between stages.
Spark is implemented on the Java Virtual Machine (JVM) and uses TCP/IP for inter-process communication.

This simple structure makes Spark an easily-learned framework, but it negatively impacts its performance. 
In particular, the constraints on the inter-process communication patterns, the lack of mutable and more flexible distributed datatypes, and the centralized scheduling of tasks, all lead to unacceptably large overheads in linear algebraic applications such as those commonly-used in machine learning~\cite{Gittens2016}.

This is the motivation for the development of Alchemist~\cite{alchemist2018}. 
Alchemist links to MPI-based libraries dynamically, and it supplies a user-friendly and efficient interface for transferring the required data from Spark executors across the network to the MPI processes, running the MPI code, and then returning the results back to the Spark application. 
Alchemist thus enables users to work within the \emph{high productivity} context of Spark, while offloading performance sensitive computations onto \emph{high performance} MPI codes. 
This approach combines Spark's strengths, such as its fast development cycle, the large collection of data analysis routines that can be used before and after the linear algebra computations, and Spark's I/O routines, while bypassing many of the inefficiencies in the Spark workflow related to its significant overheads.

The NERSC supercomputer Cori Phase 1, a Cray XC40 located at Lawrence Berkeley National Laboratory in Berkeley, California, has been used extensively in the development of Alchemist.

\subsection{Related Work}

There are previous attempts to address Spark's performance issues by either directly interfacing Spark with MPI or reimplementing Spark-like frameworks with MPI-style communication primitives.

The Smart system falls in the latter category, as a C++ reimplementation of a MapReduce-like framework built on MPI~\cite{Wang2015}. The subsequent Smart-MLlib project interfaces Spark with Smart and provides a partial reimplementation of Spark's MLlib library~\cite{Siegal2016}. Smart-MLlib facilitates data transfer between the Spark  executors and MPI processes by writing and reading from single intermediary files. This system has not been shown to be scalable and appears to no longer be under development.

The Spark-MPI project adds an MPI layer on top of Spark's standard driver-executor model~\cite{Malitsky2017}. In Spark-MPI codes, as in standard Spark codes, data is distributed as RDDs and the driver is responsible for scheduling and launching computational stages, but within stages, the executors can communicate with each other using MPI primitives. While promising, the project is still in its early stages, and it is unclear if or how it is possible to call existing MPI-based libraries from within Spark. 

The MPIgnite framework similarly extends the
Spark programming model by providing access to MPI peer-to-peer and collective communication primitives~\cite{Morris2017}. This project is also in its early stages, and no performance results were provided. 
Importantly, both Spark-MPI and MPIgnite deliberately diverge from the MapReduce programming model of Spark, so codes written for these systems are not portable to standard Spark installations.

The project closest in spirit to our approach (in that it facilitates invoking existing MPI-based libraries from Spark) is Spark+MPI~\cite{Anderson2017}. Under this framework, data is serialized and transferred from Spark to an existing MPI-based library using RAM disks, and the output is written to disk using the fault-tolerant HDFS format, which is read by Spark and loaded back into an RDD. Spark+MPI offers a simple API and supports sparse data formats.
Further, it is clear that existing codebases can be updated to call more efficient implementations of computational primitives in Spark+MPI with essentially line edits. However, the reported overheads of this system are significant when compared to the compute time, and the data sets used were relatively small, so it is not yet clear if this approach is scalable or appropriate for large-scale (in particular, dense) data sets.

Alchemist differs in important ways from these previous systems, most notably in the following ways:
\begin{itemize}
\item Alchemist does not change the Spark programming model: each call to an
MPI routine constitutes a separate stage of a Spark job. In particular, existing codebases can be line-edited to replace inefficient computations with calls to efficient MPI-based implementations.  
\item Alchemist uses sockets to transfer information between Spark and MPI.
This avoids incurring large slow-downs due to disk IO, and it similarly avoids the prohibitive memory overheads that can arise when RAM disk methods are used to transfer large data sets.
\item Alchemist is explicitly engineered for the transfer of large, dense matrices, which are worst-case use cases for some other Spark-MPI bridges. 
\item Alchemist has a flexible and extensible interface: any MPI-code that allows the user to specify the MPI communicator can be wrapped. Users can either write their performance critical codes in MPI and wrap them with Alchemist, or find pre-written codes and wrap them with Alchemist.
\end{itemize}

\begin{figure*}
\includegraphics[width=\textwidth]{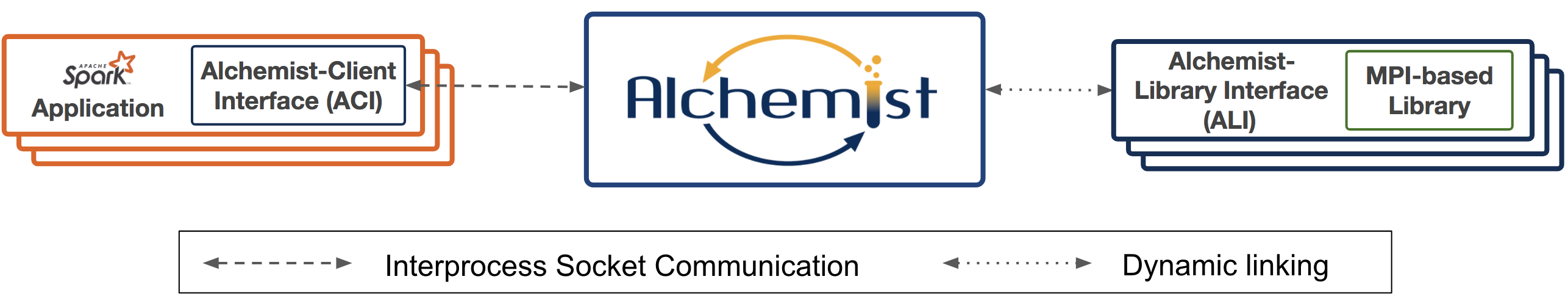}
\caption{Outline of the Alchemist framework. One or more Spark applications can each call one or more MPI-based libraries to help accelerate some numerical computations. Alchemist acts as an interface between the applications and the libraries, with data between Alchemist and the applications transmitted using TCP/IP sockets. Alchemist loads the libraries dynamically at runtime.}
\label{alchemist_architecture}
\end{figure*}

\subsection{Outline}

The rest of this paper is organized as follows. 
In Section \ref{sec:design}, we give an extensive discussion of the design and implementation of Alchemist, as well as the motivation for some of the design choices that have been made. 
In Section \ref{sec:usage}, we discuss some points related to the usage of Alchemist, with special emphasis on how to use it on Cori. 
The results of some illustrative experiments are shown in Section \ref{sec:experiments}.
(We encourage interested readers to see our original paper which introduced Alchemist~\cite{Gittens2018a} for complementary experiments, with an emphasis on relevant data science applications, including truncated PCA/SVD computations on even larger data sets of sizes up to 17.6TB.) 
In Section \ref{sec:conclusion}, we conclude with some final thoughts and future work.

\section{Design}
\label{sec:design}

This section describes the design and implementation of Alchemist. The Alchemist framework consists of the core Alchemist system and two interfaces: an Alchemist-Client Interface (ACI) and an Alchemist-Library Interface (ALI). See Figure~\ref{alchemist_architecture} for an illustration. Spark applications import the ACI, which provides functions to interact with Alchemist and to gain access to the MPI routines exposed through Alchemist. At present, all communication from the Spark application to Alchemist occurs through the ACI via TCP/IP sockets.

Alchemist, in turn, calls the desired MPI-based library through the associated ALI. Every MPI-based library has an associated ALI that consists of a thin wrapper around the MPI code whose functionality is to be exposed. The ALIs import their associated MPI libraries and need to be compiled as dynamic libraries. Alchemist then loads every ALI that is required by some Spark application dynamically at runtime.

Alchemist connects these two interfaces by providing a structure for data interchange. In the remainder of this section, we give a more detailed discussion of the inner workings of the Alchemist framework.

\subsection{The Alchemist-Client Interface}
\label{subsec:transmitting_data}

Alchemist has a server-based architecture, with one or more Spark applications being able to connect to it concurrently, assuming sufficient Alchemist workers are available. Each Spark application has to import the ACI and start an \texttt{AlchemistContext} object in order to establish a connection with Alchemist (more details on this are presented in the next section).

We broadly distinguish between two different kinds of input and output parameters: those that require distributed data structures for their storage; and those that do not. In the former category, we have, for instance, distributed dense matrices of floating point numbers. These structures must be transmitted between the Spark executors and Alchemist workers. In the latter category, we include parameters such as step sizes, maximum iteration counts, cut-off values, etc.. Such parameters are transferred easily from the application to Alchemist by using serialization, and they require communication only between the Spark and Alchemist drivers.

The critical functionality of Alchemist is an efficient implementation of communication for distributed data structures. In particular, in order to serve as a bridge, it must support sending distributed input data sets from the application to the library, as well as returning distributed output data sets (if any) from the library to the application. The design goals of the Alchemist system include making the transmission of these distributed data sets easy to use, efficient, and scalable.

Generally, there are three approaches that could be used to transmit the data:
\begin{itemize}
\item File I/O. Writing the distributed data set to a distributed file format on one side, and then reading it on the other side, has the benefit of being easy to use and, if HDFS is used, fault-tolerant. This approach will generally tend to be very slow when working with standard HDDs, although it has been argued that using an array of SSDs as the storage platform would alleviate this problem. However, SSDs are an expensive investment, and thus there is generally insufficient availability on supercomputers or in cluster centers to be useful to users with large data sets.
\item In-memory intermediary. A second option is to use an intermediate form of storage of the data in memory that can be read by both sides. This could be done using shared memory (commonly available on Linux systems), or in-memory distributed storage systems such as the open-source Apache Ignite or Alluxio. Unfortunately, we already need two copies of the data, one in the application, and the other in Alchemist, and since we are considering very large data sets, having a third copy of the data in memory is undesirable. Nevertheless, this may be an attractive option under certain circumstances.
\item Data transfer from the Spark processes directly to the MPI processes using sockets. This third option is an in-memory procedure and is therefore very fast, while not requiring an additional copy of the data set. This is therefore the most practical option for our purposes.
\end{itemize}

Alchemist and the ACI open multiple TCP/IP sockets between the Spark executors and Alchemist workers for communication, as well as one socket connection between the two driver processes. The communication is asynchronous, allowing not only multiple Spark drivers to connect to the Alchemist driver concurrently, but also to accommodate the case where each Alchemist worker receives data from several Spark executors. 
See Figure \ref{alchemist_communication} for an illustrative example and Section~\ref{subsec:example} for the associated discussion. 
Network communication in Alchemist is implemented using the Boost.Asio library~\cite{boostasio2018}. 

RDDs store their contents in rows. When transferring the data from a partition to a recipient Alchemist worker, the Spark executor sends each row of the RDD partitions to the recipient worker by transmitting the row as sequences of bytes. The received data is then recast to floating point numbers by the worker. Conversely, the transmission of matrices from the Alchemist workers to the application is done in a similar row-wise fashion. In the applications we consider in this paper, the distributed data sets are dense matrices of floating point numbers, which means that Alchemist currently sends and receives data using Spark's {\tt IndexedRowMatrix} RDD data structure.

As the Alchemist workers receive the data from the Spark executors, they store it in a distributed matrix using the Elemental library, which is described next. 

\subsection{The Elemental library}
\label{subsec:elemental}

Alchemist makes use of Elemental~\cite{elemental2017}, an open-source software package for distributed-memory dense and sparse-direct linear algebra and optimization. Elemental provides a convenient interface for handling distributed matrices with its {\tt DistMatrix} class, which is what Alchemist uses to store the data being transmitted from the RDDs. As an added benefit, Elemental provides a large suite of sequential and distributed-memory linear algebra operations that can be used to easily manipulate the distributed matrices. Copying data from distributed data sets in Spark to distributed matrices in Elemental requires some changes in the layout of the data, a task that is handled by Alchemist. The Elemental distributed matrix then serves as input to the C+MPI routines in the MPI-based library that Alchemist calls. 

Although Alchemist at present only directly supports MPI-based libraries that make use of Elemental, it is nonetheless possible to use MPI-based libraries built on top of other distributed linear algebra packages, for instance ScaLAPACK and PLAPACK. However, at present, this requires the use of wrapper functions to convert Elemental's distributed matrices to the appropriate format, and this will incur additional overhead, including potentially an additional copy of the data in memory. Support for certain other distributed linear algebra packages will be added in the future. 

\subsection{The Alchemist-Library Interface}
\label{subsec:ali}

\begin{figure*}
\includegraphics[width=\textwidth]{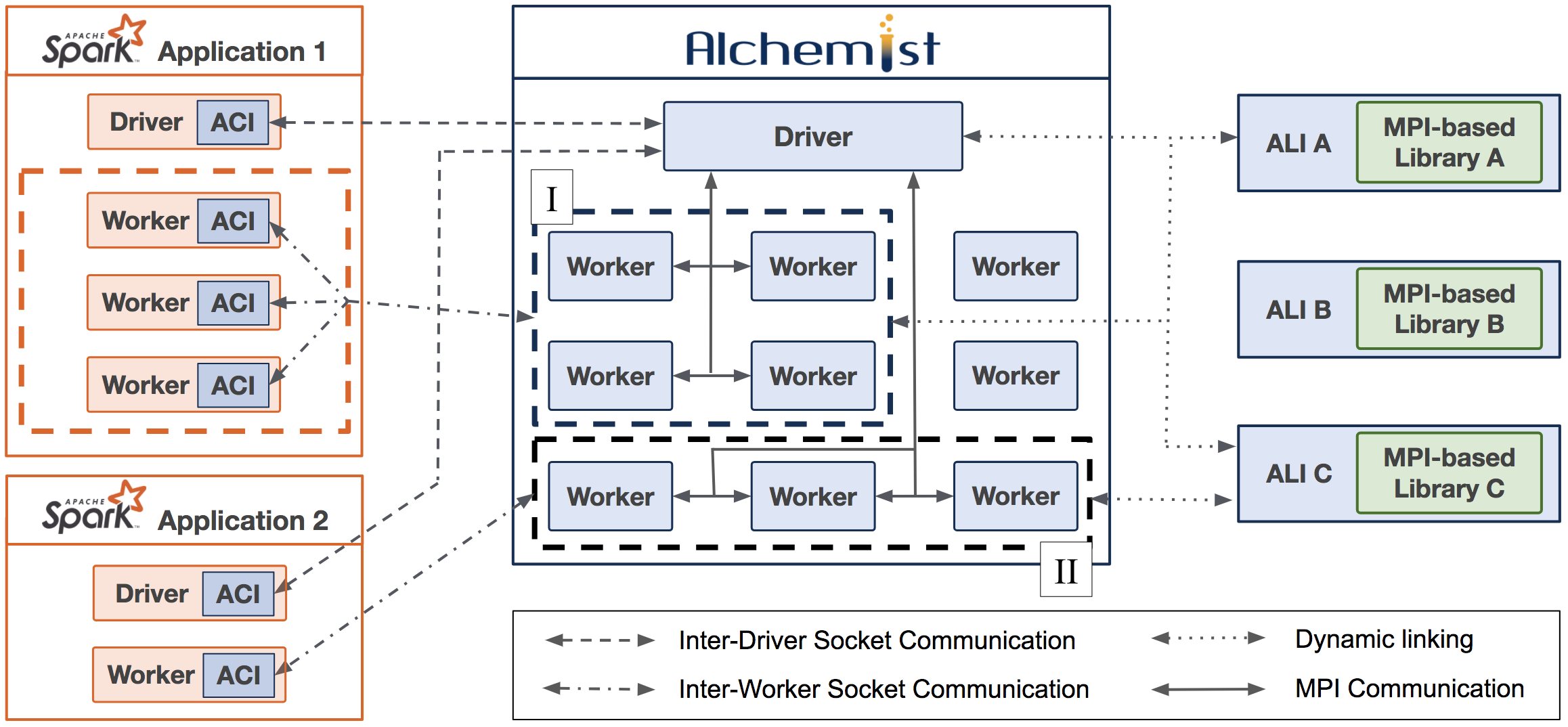}
\caption{An illustration of the Alchemist framework. See Section~\ref{subsec:example} for an explanation.}
\label{alchemist_communication}
\end{figure*}

After starting the 
\texttt{AlchemistContext} object, the Spark application should let Alchemist know which libraries it wishes to use, as well as the location of the associated ALIs for those libraries. An ALI is a shared object (i.e. a dynamic library) written in C/C++ that Alchemist can link to dynamically at runtime, assuming that there is a Spark application that wants to use it.

ALIs are necessary because Alchemist itself does not have direct knowledge of the MPI-based libraries. Each ALI imports its MPI-based library and provides a generic interface that Alchemist uses to provide it with the name of the routine in the library, and to send and receive input and output parameters to and from the routine in a pre-specified format.

The basic workflow is as follows:
\begin{itemize}
\item The Alchemist workers receive the distributed data and store it in one or more Elemental \texttt{DistMatrices}.
\item The Alchemist driver receives all the necessary metadata and non-distributed input parameters from the Spark application, including the name of the library and the routine in it that the application wants to use.
\item Alchemist calls the ALI and passes on the name of the routine, the non-distributed input parameters, and pointers to the \texttt{DistMatrices}.
\item The ALI then calls the routine in the MPI-based library with the supplied input parameters in a format appropriate for that specific library, and it returns the result in a format appropriate for Alchemist once the computations have completed. 
\item Alchemist passes the results to the Spark application via the ACI similarly to before, i.e., distributed matrices between the workers and non-distributed data between the drivers.
\end{itemize}

The use of an interface that gets linked to dynamically at runtime keeps Alchemist flexible by avoiding  the need to maintain a wrapper function inside Alchemist for every function in each MPI-based library that an application may want to call. Such a centralized system would incur a significant amount of work for each additional library that gets added, it would not give users the option to add libraries in a portable manner, and it would impose a severe maintenance and support burden.

\subsection{An Illustrative Example}
\label{subsec:example}

Figure~\ref{alchemist_communication} shows an example of the Alchemist framework at work. Alchemist is running on ten processes, one of which is the driver. Spark Application 1 has four processes, three of which are executors. The Spark driver connects to the Alchemist driver using the ACI and requests four of the available Alchemist workers, as well as access to the MPI-based libraries A and C. Alchemist loads the libraries and gives the Spark application access to four of its workers (group I in the diagram), which the ACI in each Spark process then connects to. 

Metadata and non-distributed data is sent between the drivers, and distributed data is sent from the Spark workers to Alchemist's worker group I, where the data is stored as an Elemental distributed matrix. Each worker in group I calls the required functions in libraries A and C through the associated ALIs and lets the libraries do the necessary computations.

In the meantime, a second Spark application, with just one executor, connects to Alchemist and requests access to three Alchemist worker nodes and library C. Alchemist grants access to three of its remaining workers (group II) and the Spark worker connects to them. After transferring all the data, the desired functions in library C are called. After completion, the output of all the computations is sent from the Alchemist workers to the connected Spark executors. In this example, no application made use of library B, so there was no reason for Alchemist to dynamically link to it.

The current architecture used by Alchemist, where the Spark application and the core Alchemist system run on separate groups of nodes, is due to Spark and Alchemist presently not being able to run on the same nodes on Cori. The specific reasons for this are not yet clear and future releases of Alchemist may not be subject to this requirement.
\section{Using Alchemist}
\label{sec:usage}

\lstset{language=Scala,
  basicstyle=\small\ttfamily,
  breaklines=true,
  }

Alchemist is designed to be easily deployable, flexible, and easy to use. The only required import in a Spark application is the ACI, and the ACI works with standard installations of Spark.

In this section, we discuss various aspects of using Alchemist: 
its dependencies and how to start it are discussed in Sections~\ref{subsec:dependencies} and~\ref{subsec:starting}, respectively; 
the API is presented in Section~\ref{subsec:api};
using library wrappers is discussed in Section~\ref{sxn:using_lib_wrappers}; 
and the basic procedure for adding a new MPI-based library to the system is illustrated in Section~\ref{subsec:adding_library}.

\subsection{Dependencies}
\label{subsec:dependencies}

Alchemist is written in C++11 and currently has the following dependencies:
\begin{itemize}
\item Any common implementation of MPI 3.0 or higher~\cite{mpi2015}, such as recent versions of Open MPI~\cite{openmpi2018} or MPICH~\cite{mpich2018} (or its variants).
\item The \textit{Boost.Asio} library~\cite{boostasio2018} for communicating with the ACI over sockets, as mentioned above. 
\item The \textit{Elemental} library~\cite{elemental2017} discussed previously for storing the distributed data and providing some linear algebra operations.
\item The \textit{spdlog} library~\cite{spdlog2018}, a fast, thread-safe C++ logging library for displaying Alchemist's output.
\end{itemize}
In addition, each MPI-based library (and its dependencies) that the Spark application wants to use should be installed on the system.

\subsection{Starting Alchemist}
\label{subsec:starting}

The procedure for starting Alchemist varies slightly depending on the system that one is working on, but the overall approach remains the same and is summarized in the following steps: 
\begin{enumerate}
\item The user starts Alchemist and specifies the number of MPI processes. Similarly to Spark, one Alchemist process acts as the driver and all others are worker processes. 
\item The Alchemist driver provides the IP address of the node and port number that it is running on, and the Spark driver then connects to the driver via the ACI.
\item The Spark driver can then request a user-defined number of Alchemist workers. Alchemist allocates the workers to the application (assuming a sufficient number of workers is available) and sends their IP addresses and the ports that they are running on to the ACI. The ACI then facilitates the TCP/IP socket connections of the Spark workers with the allocated Alchemist workers.
\end{enumerate}

Communication between the drivers and workers on either side can begin once all the sockets have been opened.

The Alchemist driver process receives control commands from the Spark driver, and it relays the relevant information to the worker processes.
This communication is enabled by a dedicated MPI communicator for each connected Spark application, where the communicator includes the Alchemist driver and all workers allocated to that application. 

On Cori Phase 1, Alchemist is started by running the script \texttt{Cori-start-alchemist.sh}, which needs to be run after a batch/interactive job has been launched. The script takes the number of nodes that Alchemist needs to run on and the number of cores per Alchemist worker (and driver). The remainder of the nodes allocated to the Cori job are then available for use by the Spark applications. The driver outputs its hostname, IP address and port number to disk, where it can be read in by the Spark driver's ACI.

Allowing flexibility in the number of nodes running Spark and Alchemist processes enables more resources to be allocated to Spark for jobs where it will do significant computing, and more resources can be allotted to Alchemist otherwise. We note that the number of MPI processes tends to be more than the number of Spark executors, because MPI jobs typically benefit from much higher levels of parallelism than Spark.

\subsection{The API}
\label{subsec:api}

The API of Alchemist, as far as the user is concerned, is restricted to the interaction of the Spark application with the ACI. The required syntax to get started simply involves importing the ACI, creating a new \texttt{AlchemistContext} object, and loading one or more libraries. If \texttt{sc} is an existing \texttt{SparkContext} instance, a sample code looks as follows:

\begin{lstlisting}
import alchemist.{Alchemist, AlMatrix}

// other code here ...

val ac = new Alchemist.AlchemistContext(sc, numWorkers)

// maybe other code here ...

ac.registerLibrary(ALIlibAName, ALIlibALocation)
\end{lstlisting}
Here, the hypothetical ALI {\tt ALIlibA} for the hypothetical MPI library libA has name {\tt ALIlibAName} and is located at {\tt ALIlibALocation} on the file system.

Alchemist uses matrix handles in the form of {\tt AlMatrix} objects, which act as proxies for the distributed data sets stored on Alchemist. After transmitting the data in an RDD to Alchemist, Alchemist returns an {\tt AlMatrix} object, which contains a unique ID identifying the matrix to Alchemist, as well as other information such as the dimensions of the matrix.

Similarly, for every output matrix that an MPI-based routine creates, an output matrix returns an {\tt AlMatrix} object to the application. These {\tt AlMatrix} objects allow the user to pass the distributed matrices within Alchemist from one library function to the next. Only when the user explicitly converts this object into an RDD will the data in the matrix be sent between Alchemist to Spark. In this way the amount of data transferred between Alchemist and Spark is minimized.  

If {\tt A} is an {\tt IndexedRowMatrix} in the application, and the function {\tt condest} in libA estimates the condition number of its input matrix, then the following sample code shows how to call the routine from Spark:
\begin{lstlisting}
val alA = AlMatrix(A)

val output = ac.run(ALIlibAName, "condest", alA)
\end{lstlisting}
The {\tt ac.run} function takes a variable length argument list, where the first argument is the name of the MPI-based library to use, the second is the name of the routine in the library that is being called, and the rest are the input parameters (in the above example there was just one input, {\tt AlA}). The output parameters are stored in the parameters list {\tt output}. See the documentation~\cite{alchemist2018} for more complicated examples.

After all the MPI computations have been completed and the
output {\tt AlMatrix} objects have been retrieved to {\tt IndexedRowMatrix} objects, the user can stop the {\tt AlchemistContext} instance using
\begin{lstlisting}
ac.stop()
\end{lstlisting}
similarly to how an instance of {\tt SparkContext} is stopped. 

Note that the API may be tweaked in future releases.

\subsection{Using library wrappers}
\label{sxn:using_lib_wrappers}

The API can be simplified even more by creating \textit{library wrappers}. For instance, to create a library wrapper for {\tt ALIlibA} from the previous subsection, we create a new Scala package, {\tt libA} say. Each routine in libA that will be called can then be given its own object that takes the input parameters as arguments. For the {\tt condest} routine, for instance, we could have:
\begin{lstlisting}
package alchemist.libA

import alchemist.{Alchemist, AlMatrix}

// other code here ...

object CondEst {
  def apply(alA: AlMatrix): Float = {
    ac.run("libA", "condest", alA)
  }
}
\end{lstlisting}
In this case, the sample Spark application above would then be modified as follows:
\begin{lstlisting}
import alchemist.{Alchemist, AlMatrix}
import alchemist.libA.CondEst

// other code here ...

val ac = new Alchemist.AlchemistContext(sc, numWorkers)

ac.registerLibrary("libA", ALIlibALocation)

// maybe other code here ...

val alA = AlMatrix(A)
val condNum = CondEst(alA)
\end{lstlisting}
As mentioned above, library wrappers for MPI-based libraries provide a method for giving the user a simple API for using the libraries, even if the libraries themselves have a complicated syntax; and one benefit of this approach is that one can easily mimic the API used by, for instance, MLlib. This way, one would have to only make minimal changes to existing code when switching from MLlib for some computation to an MPI-based library called through Alchemist.

\subsection{Adding a new library}
\label{subsec:adding_library}

Adding an MPI-based library to Alchemist is fairly straightforward, requiring only the implementation of the ALI for that particular library. As mentioned previously, the ALI is a shared object that Alchemist dynamically links to at runtime. The current version of Alchemist requires the library to work with Elemental distributed matrices, but support for other distributed matrix formats may be included in a future release.

A custom ALI has to implement the \texttt{Library} and \texttt{Parameters} header files that come with the Alchemist distribution. The \texttt{Parameters} header file performs the serialization and deserialization of a wide array of standard types, as well as pointers to Elemental distributed matrices. Input and output parameters are transferred between Alchemist and the ALI using the functions defined in this header.

The \texttt{Library} header declares a handful of virtual functions that need to be implemented by the ALI. In particular, the \texttt{run} function takes the name of the desired function in the library and arrays of input and output parameters. It is up to this function to call the appropriate function in the library with the supplied input parameters, and then store the return parameters appropriately.

Once an ALI has been created for a particular library, it can be distributed along with Alchemist so that other users can also benefit from it, in case they need to use the same library. We also encourage developers of ALIs to provide library wrappers that can be used directly within the Spark application, as discussed in the previous subsection. 

\section{Experiments}
\label{sec:experiments}

Having described the Alchemist framework, in this section we illustrate how much more effective Spark is at some standard linear algebra operations when used in conjunction with Alchemist rather than when used alone. 

A basic operation that underlies almost all of numerical linear algebra (and which illustrates certain fundamental challenges for Spark) is matrix multiplication, which we look at in Section~\ref{subsec:mat_mult}. 
Then, in Section~\ref{subsec:truncated_svd}, we compare timing results for the truncated singular-value decomposition (SVD). 
In general, the main overhead when using Alchemist is the time it takes to transfer data between the Spark application and Alchemist.
Thus, in Section~\ref{subsec:comm_times}, we look at some timing results for transferring a 400GB distributed matrix. 
Additional results using larger data sets can be found in our recent paper which introduced Alchemist~\cite{Gittens2018a}.

All of these experiments were run on Cori Phase 1 (also known as the Cori Data Partition system), which has two 2.3 GHz 16-core Intel Haswell processors on each of its 2,388 nodes, and 128 GB of memory per~node.

\subsection{Matrix multiplication}
\label{subsec:mat_mult}

\begin{table*}
\centering
\begin{tabular}{ c | c | c || c | c || c | c | c || c }
    \hline
    \multicolumn{3}{ c||}{Matrix dims.} & Size of  & Number & \multicolumn{3}{ c||}{Alchemist times (s)} & Spark compute  \\
    \,\,m\,\, & \,\,n\,\, & k & result (GB) & of nodes & Send & Compute & Receive & time (s) \\
    \hline
    10 & 10 & 10 & 0.8 & 1 & 5.9 & 6.6 & 2.2 & 160.3 \\
    50 & 10 & 30 & 12 & 1 & 16.7 & 75.7 & 19.4 & 809.3 \\
    100 & 10 & 70 & 56 & 2 & 32.5 & 178.7 & 55.8 & NA (476.7s) \\
    300 & 10 & 60 & 144 & 4 & 69.4 & 171.7 & 66.8 & NA (1127.3s) \\
    \hline
\end{tabular}
  \caption{Computation and Communication times for Matrix Multiplication in Spark and Spark+Alchemist. In the last two experiments, Spark failed after 476.7 and 1127.3 seconds, respectively.}
  \label{table:mat_mult}
\end{table*}

Spark's sole distributed datatype, the RDD, is a distributed array and therefore large matrices are most naturally represented in Spark as RDDs of either rows or columns. This poses a significant challenge for applications where two matrices must be multiplied: if the two matrices are both dense and row-distributed, communication-efficient matrix multiplication requires one of them to be relaid out in a column-distributed format. Transposing a dense $n \times n$ row-distributed matrix $A$ is accomplished by exploding the matrix into an RDD with $n^2$ rows of the form $(i, j, A[i,j])$, and then collecting this RDD back into an RDD of the columns of $A$. This operation is costly in terms of both memory usage, since RDDs are immutable, and communication, since it involves an all-to-all shuffle. 

Partly for this reason, there is no direct implementation of matrix multiplication in Spark's standard linear algebra provided by MLlib. Instead, one must first convert from {\tt IndexedRowMatrix} objects to {\tt BlockMatrix} objects, to use the latter's implementation of block matrix multiplication: 
\begin{lstlisting}
val A : IndexedRowMatrix = ...
val B : IndexedRowMatrix = ...
val C : IndexedRowMatrix = A.toBlockMatrix().multiply(B.toBlockMatrix()).toIndexedRowMatrix
\end{lstlisting}
This {\tt IndexedRowMatrix} to {\tt BlockMatrix} conversion is accomplished using the same explosion and collection process described above. 

We used Alchemist to wrap the Elemental matrix multiplication routine (GEMM), and we compared the time costs of using Alchemist for matrix multiplication versus that of using just Spark. 

In this experiment, we multiplied two matrices of dimensions $m \times n$ and $n \times k$, respectively. The matrices were random dense matrices generated within Spark. 
(The runtimes for this operation are independent of the matrix properties, so random matrices should be regarded as representative.) 
The multiplication was performed both in Spark and in Alchemist, on the same number of nodes for both. See Table~\ref{table:mat_mult} for the timing results. There is some variability in the times when performing the experiments, so three runs were performed for each configuration and the reported times shown in Table~\ref{table:mat_mult} are average results.

The first two cases (i.e., the first two rows) show that even if the matrices fit on one machine, one still benefits from using MPI for the multiplication because of the improved timing results. The cases where the matrices are distributed over more than one node illustrate how delicate shuffling between machines is in Spark. While we are not claiming that it will fail on every occasion, it did for all the runs we performed, and the fact that Spark explodes the matrices into (i,j,k) pairs to do matrix multiplication makes multi-machine matrix multiplies unreliable.

We should note that, for these computations we used Spark's default values for blocking and parallelism. Improved results may be obtained by finding an optimal configuration for a given pair of matrices, but this is unnecessary when using MPI-based routines, and the timing results are unlikely to be competitive in any case.

\subsection{Truncated SVD}
\label{subsec:truncated_svd}

Principal component analysis (PCA) is a ubiquitous procedure used to reduce the dimensionality of large data sets and to uncover meaningful structure in data. The computational primitive underlying PCA is the SVD. To demonstrate the advantage of using Alchemist to compute SVDs of medium-to-large data sets, we compute the rank-$k$ truncated SVD (i.e. we compute only the first $k$ singular values, for some modest value of $k$) of randomly generated double-precision dense matrices with dimensions $m \times 10,000$, where we let $m$ be $312,500$, $625,000$, $1,250,000$, $2,500,000$, and $5,000,000$. This yields matrices ranging in size from 25GB to 400GB.

We wrote our own MPI-based implementation of the truncated SVD using ARPACK~\cite{ARPACK1997} and Elemental. In Spark, we use the MLlib implementation of SVD, which similarly uses ARPACK, by calling the {\tt IndexedRowMatrix} class's {\tt computeSVD} method. We compute the first 20 singular values and vectors of the matrices using 22 Spark nodes, with one executor per node and using default parallelism, and 8 nodes of Alchemist processes, with 16 Alchemist workers per node. Jobs were run on Cori's debug queue, so they were limited to a maximum of 30 minutes (1800 seconds). 

\begin{figure}[htb!]
\centering
\subfloat[ ]{\includegraphics[width=3.2in]{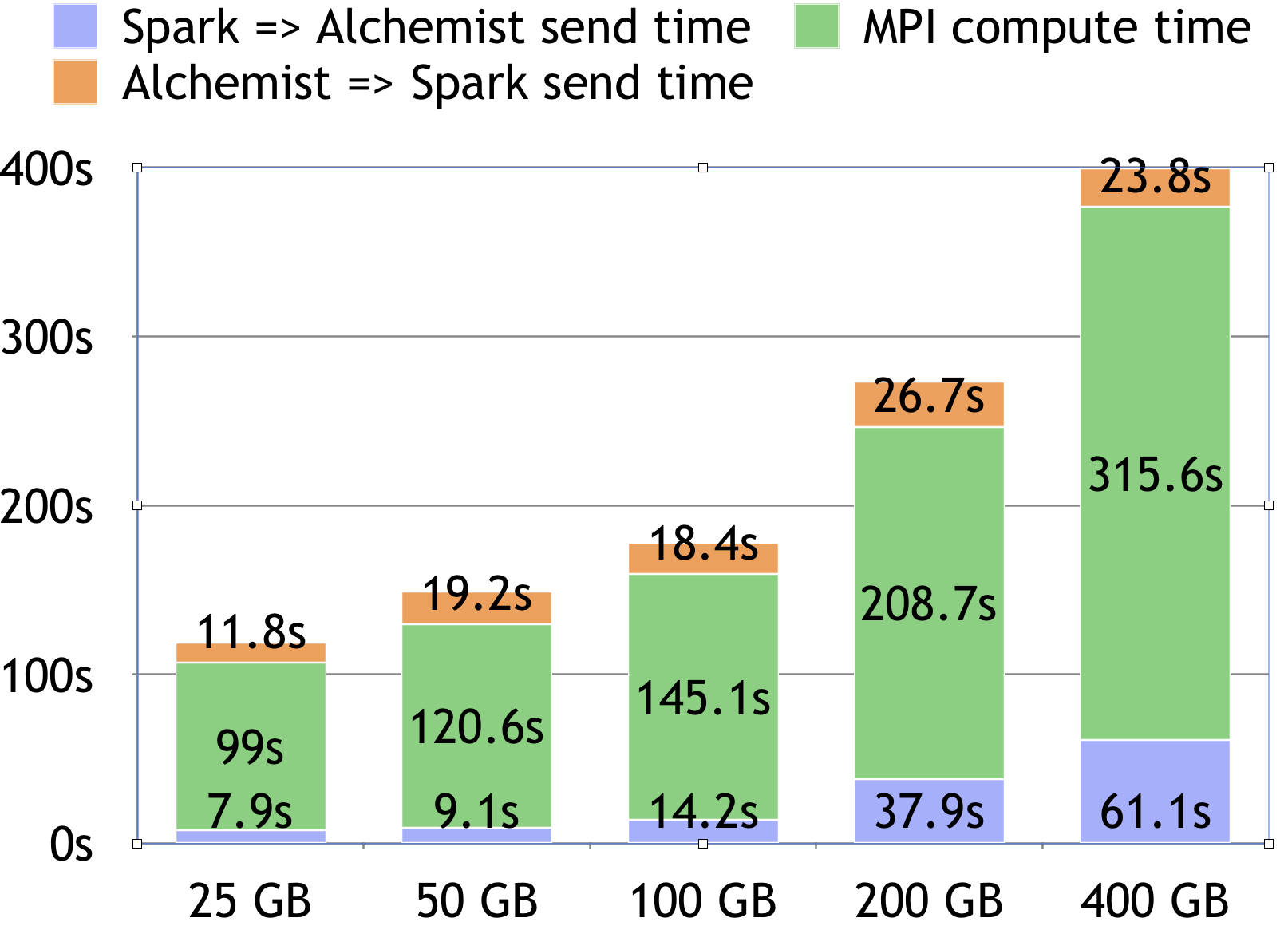}}
\caption{Alchemist's overheads are due to the time it takes to transmit the distributed data sets between Spark and Alchemist. While non-negligible, the overheads constitute just 20\% of the overall running time of the truncated SVD procedure. This is a significant improvement on the overheads incurred by Spark (see the discussion in~\cite{Gittens2016}).}
\label{fig:SVD_alchemist_overheads}
\end{figure}

\begin{figure}[htp!]
\centering
\subfloat[ ]{\includegraphics[width=3.2in]{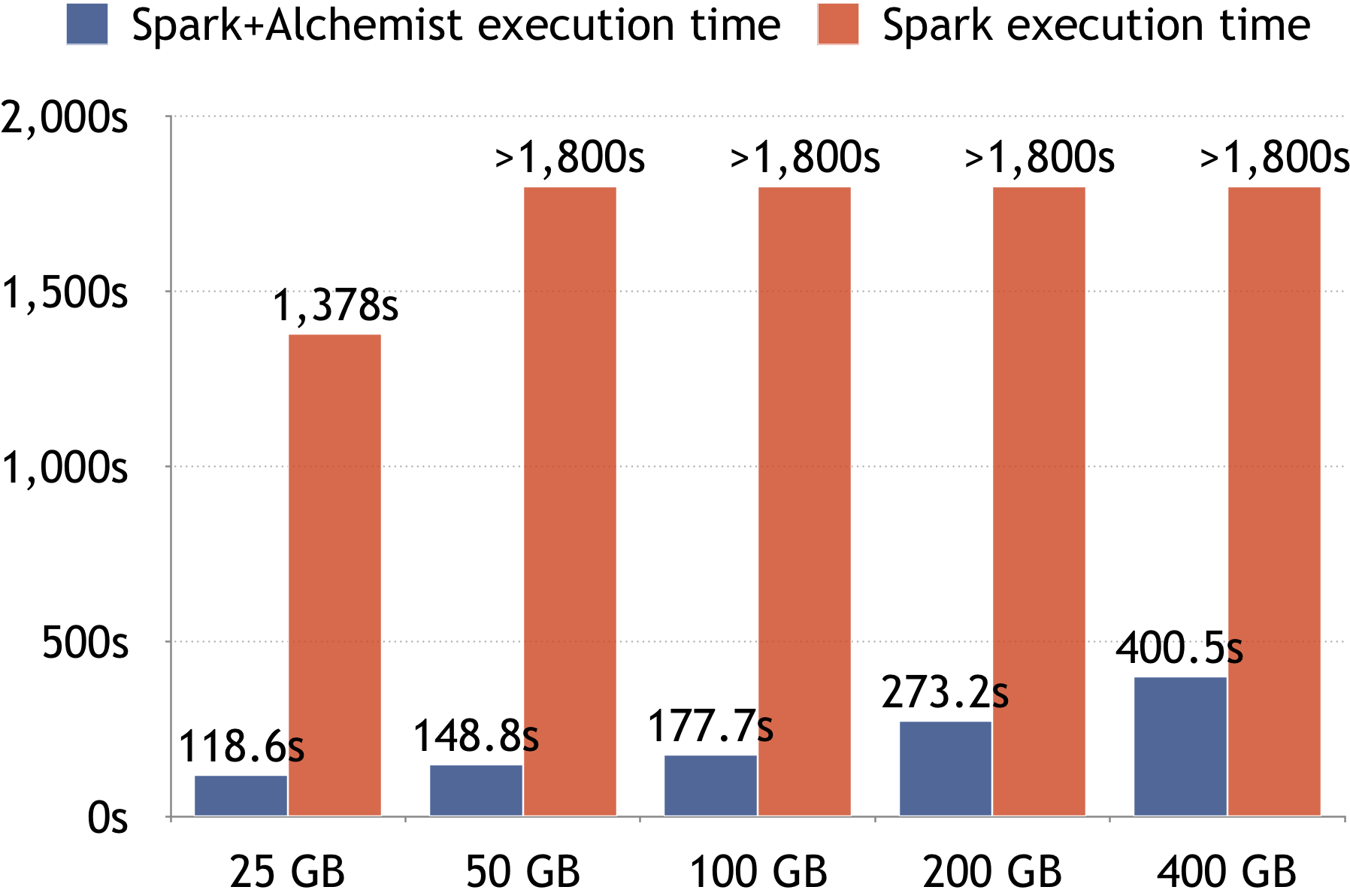}}
\caption{Comparison of running times (compute times and overheads combined) between Spark and Alchemist for the truncated SVD of a randomly generated dense matrix. Clearly, Alchemist provides significantly improved performance, as the Spark computations do not complete within the 30 minute limit for all but the smallest matrix.}
\label{fig:SVD_spark_vs_alchemist}
\end{figure}

The primary Alchemist overhead is the time it takes to transfer large data sets between Spark and Alchemist. Figure~\ref{fig:SVD_alchemist_overheads} shows the time it takes to transfer the data compared to the time it takes to perform the computations. 
It is evident that the overheads take up approximately 20\% of the total runtime. Comparing this to the overheads encountered by Spark, which are orders of magnitude larger than the actual compute times (see the discussion in~\cite{Gittens2016}), we see that there is a significant improvement in performance when using Alchemist to perform the SVD computations. 
This difference is further highlighted in Figure~\ref{fig:SVD_spark_vs_alchemist}, which shows an enormous performance difference, including that Spark was unable to complete the SVD computations in the allotted time for all but the smallest matrix.

This experiment demonstrates that off-loading linear-algebraic tasks from Spark onto Alchemist can result in significant decreases in runtime, even when the overheads of the data transfer are taken into account.




\subsection{Communication times}
\label{subsec:comm_times}

As seen in the previous experiments, the most significant overhead when using Alchemist is the cost of transmitting the data. The Spark application needs to send distributed matrices to Alchemist, and then Alchemist needs to send the results of the computations back to the application. Recall that the data is distributed over the Spark executors, and each executor sends its data to one or more Alchemist workers. In this experiment we investigate how the communication times vary for a random 400GB floating point matrix, relative to the number of nodes used by the application and the number of nodes used by Alchemist. We look at the times for matrices of dimensions $5,120,000\times 10,000$ (tall and skinny) and $40,000 \times 1,280,000$ (short and wide). We look only at the times it takes to send the data from Spark to Alchemist. Recall that this is done using TCP/IP sockets.

These simulations were run in the debug queue of Cori, therefore a maximum of 64 Haswell nodes were used in total for each job. The random matrices were created within the Spark application and we let it create and store them using its default parallelism. As is to be expected when transmitting a large amount of data over any kind of network, even on a supercomputer, there is some variability in the time it takes to send a large amount of data from Spark to Alchemist, at times significant. We found that this variability is even higher on average when working in interactive mode on Cori than when running a batch job.

What is shown in the tables above is the average of three runs for each configuration, although on some occasions there were outliers where the communication seemed to stagnate, which were ignored. In Table~\ref{table:send_times_5120000_10000}, we show the time it takes to transfer a 400GB $5,120,000 \times 10,000$ matrix from the Spark executors to the Alchemist workers, for different configurations of nodes. The times are fairly consistent, with apparently a slight tendency to be minimized if the number of Spark and Alchemist nodes are the same.

\begin{table}
\centering
\begin{tabular}{ c | c  c  c  c  c  c  c }
    \hline
    \# Spark & \multicolumn{6}{ c }{\# Alchemist nodes} \\
    nodes & 8 & 16 & 24 & 32 & 40 & 48 & 56 \\
    \hline
    8 & 62.1 & 65.2 & 66.4 & 72.4 & 72.8 & 76.7 & 88.5 \\
    16 & 75.6 & 68.3 & 72.8 & 81.1 & 89.3 & 93.5 &  \\
    24 & 73.0 & 69.7 & 62.8 & 77.5 & 82.0 &  &  \\
    32 & 78.5 & 75.4 & 69.8 & 66.8 &  &  &  \\
    40 & 69.6 & 65.4 & 62.4 &  &  &  &  \\
    48 & 70.6 & 67.9 &  &  &  &  &  \\
    56 & 64.5 &  &  &  &  &  &  \\
    \hline
\end{tabular}
  \caption{Transmission times of a random 400GB matrix with dimensions $5,120,000 \times 10,000$ from a Spark application to Alchemist (in seconds). Maximum of 64 nodes in total.}
  \label{table:send_times_5120000_10000}
\end{table}

The times it takes to transfer a 400GB $40,000 \times 1,280,000$ matrix from the Spark executors to the Alchemist workers, for different configurations of nodes, is shown in Table~\ref{table:send_times_40000_1280000}. 

\begin{table}
\centering
\begin{tabular}{ c | c  c  c  c  c  c  c }
    \hline
    \# Spark & \multicolumn{6}{ c }{\# Alchemist nodes} \\
    nodes & 8 & 16 & 24 & 32 & 40 & 48 & 56 \\
    \hline
    8 & 59.8 & 50.0 & 38.5 & 30.1 & 15.2 & 14.7 & 18.6 \\
    16 & 55.8 & 34.0 & 24.5 & 20.2 & 13.9 & 13.8 &  \\
    24 & 56.2 & 34.9 & 21.9 & 18.0 & 12.4 &  &  \\
    32 & 54.1 & 30.5 & 22.1 & 15.1 &  &  &  \\
    40 & 52.9 & 30.6 & 22.7 &  &  &  &  \\
    48 & 54.5 & 27.2 &  &  &  &  &  \\
    56 & 57.6 &  &  &  &  &  &  \\
    \hline
\end{tabular}
  \caption{Transmission times of a random 400GB matrix with dimensions $40,000 \times 1,280,000$ from a Spark application to Alchemist (in seconds). Maximum of 64 nodes in total.}
  \label{table:send_times_40000_1280000}
\end{table}

There are several interesting points to make regarding the results in this table:
\begin{itemize}
\item Overall the times are less than when transferring the tall and skinny matrix.
\item While not shown here, not only were transfer times less, but they also showed far less variability than when sending the tall and skinny matrix.
\item Regardless of the number of Spark nodes, it takes more time to send data to a small number of Alchemist workers, and in fact the transfer times decrease fairly consistently as the number of Alchemist workers increases.
\end{itemize}

The reasons for the improved performance when sending short and wide matrices vs. tall and skinny ones appears to be due to how the {\tt IndexedRowMatrices} are sent to Alchemist, which is one row at a time. There are far more rows in tall and skinny matrices, so there are significantly more messages sent between the application and Alchemist. Therefore, factors such as disruptions in the network and, in particular, the current load on the network have a much larger impact on the communication times (and their variability) than when sending fewer (but longer) messages. 

It is not entirely clear why the transfer times decrease as the number of Alchemist nodes increases in the case of the short and wide matrix, but it is possibly due to limitations in how fast data can be stored in the Elemental {\tt DistMatrix} in Alchemist when the matrix is distributed over fewer nodes.

As mentioned above, at present the communication between the Spark application and Alchemist is done using TCP/IP sockets. The nodes on Cori are connected using Cray's propriety Aries interconnect, and it is expected that appreciable reductions in the average data transfer times between the ACI and Alchemist can be achieved by using a communication protocol designed for this interconnect.

\section{Conclusion}
\label{sec:conclusion}

We have described in detail the motivation for and design of Alchemist, and we have shown that calling MPI-based libraries from Spark applications using Alchemist leads to a significantly improved performance with comparatively little overhead, while still retaining the productivity benefits of working in the Spark environment. This opens the door to make the use of Apache Spark not only more efficient, but also more attractive to potential users who crave the combination of the performance of MPI with the simple interface, the productivity possibilities, and the extensive ecosystem of Spark. 

While this improvement in efficiency is to be commended, it does come at a price. Some overhead is incurred when transferring data between Spark and the MPI libraries over sockets---although this overhead is small compared to the overhead that would have been incurred by Spark, and it is less than when using the other methodologies for transferring data that were mentioned in Section \ref{subsec:transmitting_data}. Alchemist also requires two copies of the data: the RDD used by the Spark application, and the same data stored as a distributed matrix on the MPI side. This is a necessary limitation, since MPI codes are not able to access the data in the RDD directly.

Future developments for running Alchemist on Cori more efficiently include reducing data transfer times by using communication protocols designed specifically for the Cray Aries interconnect, and attempting to run Alchemist and Spark on the same nodes, which will reduce the amount of data that needs to be transferred. 

It is also important to enable Alchemist to run on 
MPI-based libraries built on top of ScaLAPACK, another framework for distributed linear algebra. The Elemental library is convenient since it allows us to make use of its user-friendly distributed matrix implementations, but other libraries will require wrapper functions to transform the data to the appropriate format. This will incur a non-negligible overhead, as well as an additional copy of the data in memory, something that should be avoided if possible. Given the longevity of ScaLAPACK, it would be beneficial to provide direct support for libraries built on top of this package. ScaLAPACK does not have the convenient interface for creating and using distributed matrices that Elemental has, but communication between the Spark executors and Alchemist processes can nonetheless work similarly to the process described in Section~\ref{sec:design}.

While we have looked at linear algebra routines in this paper, MPI-based libraries for other applications can also be used with Alchemist.
This would be particularly interesting if more machine learning libraries are available. Unfortunately, it appears that there are few MPI-based machine learning projects at the moment, and they tend to be quite limited in their scope. One exception to this is MaTEx~\cite{Matex2017}, which offers somewhat more functionality, but it does not appear to be under active development, and it is designed to be invoked as a standalone program rather than a library. Interfacing Spark with standalone MPI-based applications using Alchemist remains an open problem, since it would require either wrapper functions or file I/O to transfer the data, approaches we would like to avoid if possible. 

Finally, future versions of the system should also be able to run on cloud-based computing platforms such as AWS EC2 and Azure, and work towards this goal is currently being undertaken by making use of Docker and Kubernetes.

\section*{Acknowledgments}
  This work was partially supported by NSF, DARPA, and Cray Inc.

\bibliographystyle{plain}
\bibliography{Bibliography}

\begin{thebibliography}{10}

\bibitem{Anderson2017}
M.~Anderson, Shaden Smith, Narayanan Sundaram, Mihai Capota, Zheguang Zhao,
  Subramanya Dulloor, Nadathur Satish, and Theodore~L. Willke.
\newblock {Bridging the Gap Between HPC and Big Data Frameworks}.
\newblock In {\em Proceedings of the VLDB Endowment}, volume~10, pages
  901--912, 2017.

\bibitem{Gittens2016}
Alex Gittens, Aditya Devarakonda, Evan Racah, Michael Ringenburg, Lisa
  Gerhardt, Jey Kottalam, Jialin Liu, Kristyn Maschhoff, Shane Canon, Jatin
  Chhugani, Pramod Sharma, Jiyan Yang, James Demmel, Jim Harrell, Venkat
  Krishnamurthy, Michael~W. Mahoney, and Prabhat.
\newblock {Matrix factorizations at scale: A comparison of scientific data
  analytics in Spark and C+MPI using three case studies}.
\newblock In {\em 2016 IEEE International Conference on Big Data (Big Data)},
  pages 204--213, 2016.

\bibitem{Gittens2018a}
Alex Gittens, Kai Rothauge, Michael~W. Mahoney, Shusen Wang, Jey Kottalam,
  Prabhat, Lisa Gerhardt, Michael Ringenburg, and Kristyn Maschhoff.
\newblock {Accelerating Large-Scale Data Analysis by offloading to
  High-Performance Computing Libraries using Alchemist}.
\newblock In {\em Proceedings of the 24th ACM SIGKDD International Conference
  on Knowledge Discovery and Data Mining}, 2018.
\newblock to appear.

\bibitem{boostasio2018}
Christopher Kohlhoff.
\newblock {Boost.Asio}, 2018.
\newblock https://www.boost.org/.

\bibitem{ARPACK1997}
R.~B. Lehoucq, K.~Maschhoff, D.~C. Sorensen, and C.~Yang.
\newblock {ARPACK: Solution of Large Scale Eigenvalue Problems with Implicitly
  Restarted Arnoldi Methods}, 1997.

\bibitem{Malitsky2017}
N.~Malitsky, A.~Chaudhary, S.~Jourdain, M.~Cowan, P.~O'Leary, M.~Hanwell, and
  K.~K.~Van Dam.
\newblock {Building near-real-time processing pipelines with the Spark-MPI
  platform}.
\newblock In {\em 2017 New York Scientific Data Summit (NYSDS)}, pages 1--8,
  2017.

\bibitem{spdlog2018}
Gabi Melman.
\newblock {spdlog}, 2018.
\newblock https://github.com/gabime/spdlog.

\bibitem{mpi2015}
{Message Passing Interface Forum}.
\newblock {MPI: A Message-Passing Interface Standard: Version 3.1}.
\newblock Technical report, 2015.

\bibitem{Morris2017}
B.~L. Morris and A.~Skjellum.
\newblock {MPIgnite: An MPI-like Language for Apache Spark}, 2017.
\newblock {Poster presented at European MPI Users' Group Meeting, September
  25--28, Chicago, IL, USA.}

\bibitem{mpich2018}
{MPICH: High-Performance Portable MPI}.
\newblock https://www.mpich.org/.

\bibitem{openmpi2018}
{Open MPI: Open Source High Performance Computing}.
\newblock https://www.open-mpi.org/.

\bibitem{elemental2017}
J.~Poulson, B.~Marker, R.~van~de Geijn, J.~Hammond, and N.~Romero.
\newblock {Elemental: A new framework for distributed memory dense matrix
  computations}.
\newblock {\em ACM Transactions on Mathematical Software}, 39:1--24, 2013.

\bibitem{alchemist2018}
Kai Rothauge and A.~Gittens.
\newblock {Alchemist: An Apache Spark $<$=$>$ MPI interface}, 2018.
\newblock http://github.com/project-alchemist/.

\bibitem{Siegal2016}
D.~Siegal, J.~Guo, and G.~Agrawal.
\newblock {Smart-MLlib: A High-Performance Machine-Learning Library}.
\newblock In {\em 2016 IEEE International Conference on Cluster Computing
  (CLUSTER)}, pages 336--345, 2016.

\bibitem{Matex2017}
Abhinav Vishnu, Jeff Daily, Charles Siegel, Joseph Manzano, et~al.
\newblock {MaTEx: Machine Learning Toolkit for Extreme Scale}, 2017.
\newblock https://github.com/matex-org/matex.

\bibitem{Wang2015}
Yi~Wang, Gagan Agrawal, Tekin Bicer, and Wei Jiang.
\newblock {Smart: A MapReduce-like framework for in-situ scientific analytics}.
\newblock In {\em Proceedings of the International Conference for High
  Performance Computing, Networking, Storage and Analysis}, pages 1--12. ACM,
  2015.

\bibitem{Zaharia2016}
M.~Zaharia, Reynold~S. Xin, Patrick Wendell, Tathagata Das, Michael Armbrust,
  Ankur Dave, Xiangrui Meng, Josh Rosen, Shivaram Venkataraman, Michael~J.
  Franklin, Ali Ghodsi, Joseph Gonzalez, Scott Shenker, and Ion Stoica.
\newblock {Apache Spark: a unified engine for big data processing}.
\newblock In {\em Communications of the ACM}, volume~59, pages 56--65, 2016.

\end{thebibliography}

\end{document}